\documentstyle[preprint,psfig,amstex,epsf,aps,floats,tighten]{revtex}

\input psfig
\input epsf 
\def\ppbar{$p\overline{p}~$}             
\def\met{\mbox{${\hbox{$E$\kern-0.6em\lower-.1ex\hbox{/}}}_T~$}} 
\def\D0{D\O}                            
\begin{document}
\title{Search for Heavy Particles Decaying into Electron-Positron Pairs
    in $\boldmath p\overline p$ Collisions}
\author{                                                                      
V.M.~Abazov,$^{23}$                                                           
B.~Abbott,$^{58}$                                                             
A.~Abdesselam,$^{11}$                                                         
M.~Abolins,$^{51}$                                                            
V.~Abramov,$^{26}$                                                            
B.S.~Acharya,$^{17}$                                                          
D.L.~Adams,$^{60}$                                                            
M.~Adams,$^{38}$                                                              
S.N.~Ahmed,$^{21}$                                                            
G.D.~Alexeev,$^{23}$
% ...
\\                                                                            
\vskip 0.30cm                                                                 
\centerline{(D\O\ Collaboration)}                                             
\vskip 0.30cm                                                                 
}                                                                             
\address{                                                                     
\centerline{$^{1}$Universidad de Buenos Aires, Buenos Aires, Argentina}       
\centerline{$^{2}$LAFEX, Centro Brasileiro de Pesquisas F{\'\i}sicas,         
                  Rio de Janeiro, Brazil}                                     
\centerline{$^{3}$Universidade do Estado do Rio de Janeiro,                   
                  Rio de Janeiro, Brazil}                                     
\centerline{$^{4}$Institute of High Energy Physics, Beijing,                  
                  People's Republic of China}                                 
\centerline{$^{5}$Universidad de los Andes, Bogot\'{a}, Colombia}             
\centerline{$^{6}$Charles University, Center for Particle Physics,            
                  Prague, Czech Republic}                                     
\centerline{$^{7}$Institute of Physics, Academy of Sciences, Center           
                  for Particle Physics, Prague, Czech Republic}               
\centerline{$^{8}$Universidad San Francisco de Quito, Quito, Ecuador}         
\centerline{$^{9}$Institut des Sciences Nucl\'eaires, IN2P3-CNRS,             
                  Universite de Grenoble 1, Grenoble, France}                 
\centerline{$^{10}$CPPM, IN2P3-CNRS, Universit\'e de la M\'editerran\'ee,     
                  Marseille, France}                                          
\centerline{$^{11}$Laboratoire de l'Acc\'el\'erateur Lin\'eaire,              
                  IN2P3-CNRS, Orsay, France}                                  
\centerline{$^{12}$LPNHE, Universit\'es Paris VI and VII, IN2P3-CNRS,         
                  Paris, France}                                              
\centerline{$^{13}$DAPNIA/Service de Physique des Particules, CEA, Saclay,    
                  France}                                                     
\centerline{$^{14}$Universit{\"a}t Mainz, Institut f{\"u}r Physik,            
                  Mainz, Germany}                                             
\centerline{$^{15}$Panjab University, Chandigarh, India}                      
\centerline{$^{16}$Delhi University, Delhi, India}                            
\centerline{$^{17}$Tata Institute of Fundamental Research, Mumbai, India}     
\centerline{$^{18}$Seoul National University, Seoul, Korea}                   
\centerline{$^{19}$CINVESTAV, Mexico City, Mexico}                            
\centerline{$^{20}$FOM-Institute NIKHEF and University of                     
                  Amsterdam/NIKHEF, Amsterdam, The Netherlands}               
\centerline{$^{21}$University of Nijmegen/NIKHEF, Nijmegen, The               
                  Netherlands}                                                
\centerline{$^{22}$Institute of Nuclear Physics, Krak\'ow, Poland}            
\centerline{$^{23}$Joint Institute for Nuclear Research, Dubna, Russia}       
\centerline{$^{24}$Institute for Theoretical and Experimental Physics,        
                   Moscow, Russia}                                            
\centerline{$^{25}$Moscow State University, Moscow, Russia}                   
\centerline{$^{26}$Institute for High Energy Physics, Protvino, Russia}       
\centerline{$^{27}$Lancaster University, Lancaster, United Kingdom}           
\centerline{$^{28}$Imperial College, London, United Kingdom}                  
\centerline{$^{29}$University of Arizona, Tucson, Arizona 85721}              
\centerline{$^{30}$Lawrence Berkeley National Laboratory and University of    
                  California, Berkeley, California 94720}                     
\centerline{$^{31}$University of California, Davis, California 95616}         
\centerline{$^{32}$California State University, Fresno, California 93740}     
\centerline{$^{33}$University of California, Irvine, California 92697}        
\centerline{$^{34}$University of California, Riverside, California 92521}     
\centerline{$^{35}$Florida State University, Tallahassee, Florida 32306}      
\centerline{$^{36}$University of Hawaii, Honolulu, Hawaii 96822}              
\centerline{$^{37}$Fermi National Accelerator Laboratory, Batavia,            
                   Illinois 60510}                                            
\centerline{$^{38}$University of Illinois at Chicago, Chicago,                
                   Illinois 60607}                                            
\centerline{$^{39}$Northern Illinois University, DeKalb, Illinois 60115}      
\centerline{$^{40}$Northwestern University, Evanston, Illinois 60208}         
\centerline{$^{41}$Indiana University, Bloomington, Indiana 47405}            
\centerline{$^{42}$University of Notre Dame, Notre Dame, Indiana 46556}       
\centerline{$^{43}$Iowa State University, Ames, Iowa 50011}                   
\centerline{$^{44}$University of Kansas, Lawrence, Kansas 66045}              
\centerline{$^{45}$Kansas State University, Manhattan, Kansas 66506}          
\centerline{$^{46}$Louisiana Tech University, Ruston, Louisiana 71272}        
\centerline{$^{47}$University of Maryland, College Park, Maryland 20742}      
\centerline{$^{48}$Boston University, Boston, Massachusetts 02215}            
\centerline{$^{49}$Northeastern University, Boston, Massachusetts 02115}      
\centerline{$^{50}$University of Michigan, Ann Arbor, Michigan 48109}         
\centerline{$^{51}$Michigan State University, East Lansing, Michigan 48824}   
\centerline{$^{52}$University of Nebraska, Lincoln, Nebraska 68588}           
\centerline{$^{53}$Columbia University, New York, New York 10027}             
\centerline{$^{54}$University of Rochester, Rochester, New York 14627}        
\centerline{$^{55}$State University of New York, Stony Brook,                 
                   New York 11794}                                            
\centerline{$^{56}$Brookhaven National Laboratory, Upton, New York 11973}     
\centerline{$^{57}$Langston University, Langston, Oklahoma 73050}             
\centerline{$^{58}$University of Oklahoma, Norman, Oklahoma 73019}            
\centerline{$^{59}$Brown University, Providence, Rhode Island 02912}          
\centerline{$^{60}$University of Texas, Arlington, Texas 76019}               
\centerline{$^{61}$Texas A\&M University, College Station, Texas 77843}       
\centerline{$^{62}$Rice University, Houston, Texas 77005}                     
\centerline{$^{63}$University of Virginia, Charlottesville, Virginia 22901}   
\centerline{$^{64}$University of Washington, Seattle, Washington 98195}       
}     
\date{\today}

\maketitle

%
%
\begin{abstract}
We present results of searches for technirho
($\rho_T$), techniomega ($\omega_T$), and $Z'$ particles,
using the decay channels $\rho_T,\omega_T,Z'\rightarrow e^+ e^-$.
The search is based on 124.8 pb$^{-1}$ of data collected
by the D\O\ detector at the Fermilab Tevatron during 1992--1996.
In the absence of a signal, we set 95\% C.L. upper limits on the 
cross sections for the processes  
$p\overline p\to\rho_T,\omega_T,Z'\rightarrow e^+ e^-$ 
as a function of the mass of the decaying particle. For certain 
model parameters, we exclude the existence of degenerate $\rho_T$ and 
$\omega_T$ states with masses below about 200~GeV. We exclude a $Z'$ with mass 
below 670~GeV, assuming that it has the same couplings to fermions as the 
$Z$ boson.
\end{abstract}
\draft

\narrowtext



Historically, studies of lepton-antilepton pair production --- in particular 
$e^+e^-$ and $\mu^+\mu^-$ --- have been important discovery channels for new 
particles. The $J/\psi$, $\Upsilon$, and $Z$ resonances were all found in this 
way. Many extensions of the standard model predict the existence of particles 
that decay to lepton-antilepton pairs. Examples are heavy gauge bosons ($Z'$) 
and technihadrons ($\rho_T$, $\omega_T$). 
The lepton-antilepton signature is a preferred channel for particle searches  
in strong interactions because of the relatively low backgrounds compared to 
hadronic decay channels. Electrons and muons permit a relatively 
straightforward trigger 
and their momenta can be measured precisely. Thus particles that decay to 
$e^+e^-$ or $\mu^+\mu^-$ can be identified as resonances in the dilepton mass 
spectrum.

In this Letter, we describe a search for resonances in the dielectron mass 
spectrum in data collected by D\O\ during 1992--1996 at the Fermilab Tevatron. 
We first describe the data sample, background sources, acceptance, and 
efficiency. We then set limits on the product of the cross section and 
branching fraction for the production of such resonances and their subsequent 
decay to $e^+e^-$ as a function of the resonance mass. Finally, we compare 
this limit to predictions for hypothesized particles.


The D\O\ detector\cite{D0NIM} is a multi-purpose particle detector. It tracks 
charged particles in tracking detectors located around the interaction region. 
The energy
of particles is measured in uranium/liquid-argon calorimeters that surround
the tracking detectors. The calorimeters are housed in three cryostats. In the
central calorimeter (CC) we accept electrons with pseudorapidity 
$|\eta|<1.1$ and in the end calorimeters (EC) with $1.5<|\eta|<2.5$. 
Pseudorapidity is defined in terms of the polar angle $\theta$ relative to 
the proton beam direction as $\eta=-\ln\tan{\theta\over2}$. Electrons are 
identified as narrow showers in the electromagnetic section of the 
calorimeters, with a matching track in the drift chambers. The electron 
energy $E$ is measured with a resolution $\sigma_E$, given by 
$\left(\sigma_E/E\right)^2=\left(15\%/\sqrt{E/\mbox{GeV}}\right)^2+
\left(1\%\right)^2$. 
No distinction can be made between electrons and positrons, because the 
tracking detectors are not in a magnetic field.
 

The data sample, background predictions, event selection, 
and electron identification criteria used for this analysis are identical 
to those described in Ref. \cite{Gupta}. We require at least two electrons 
\cite{electron} with $E\sin\theta>25$ GeV. To maximize the signal efficiency, 
one electron in the CC fiducial region is not required to have a matching 
track.

\begin {figure}[htb!]
\centerline{\psfig{figure=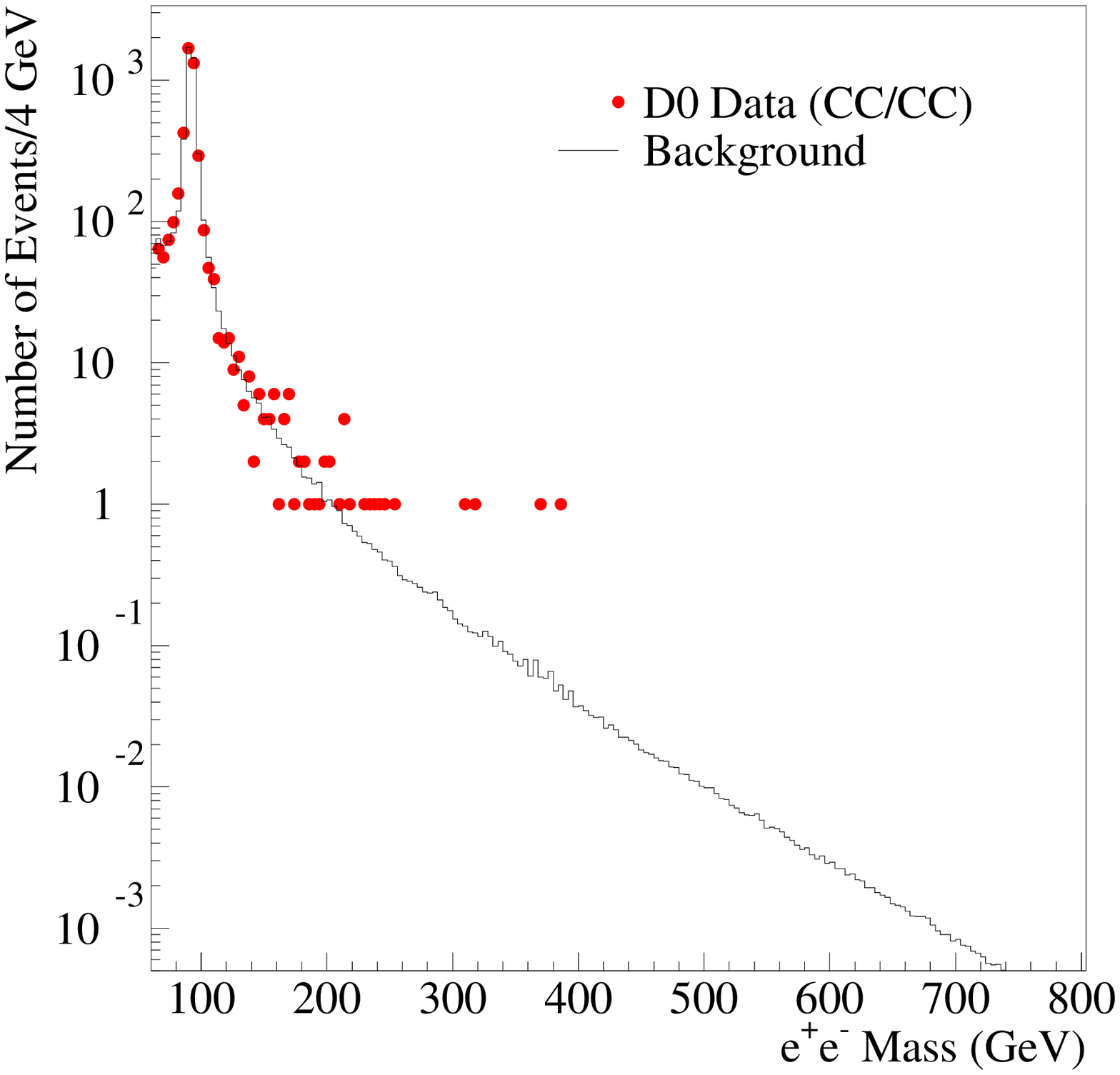,width=3.5in}}
\caption{Dielectron invariant mass spectrum for CC/CC events.}
\label{figone}

\centerline{\psfig{figure=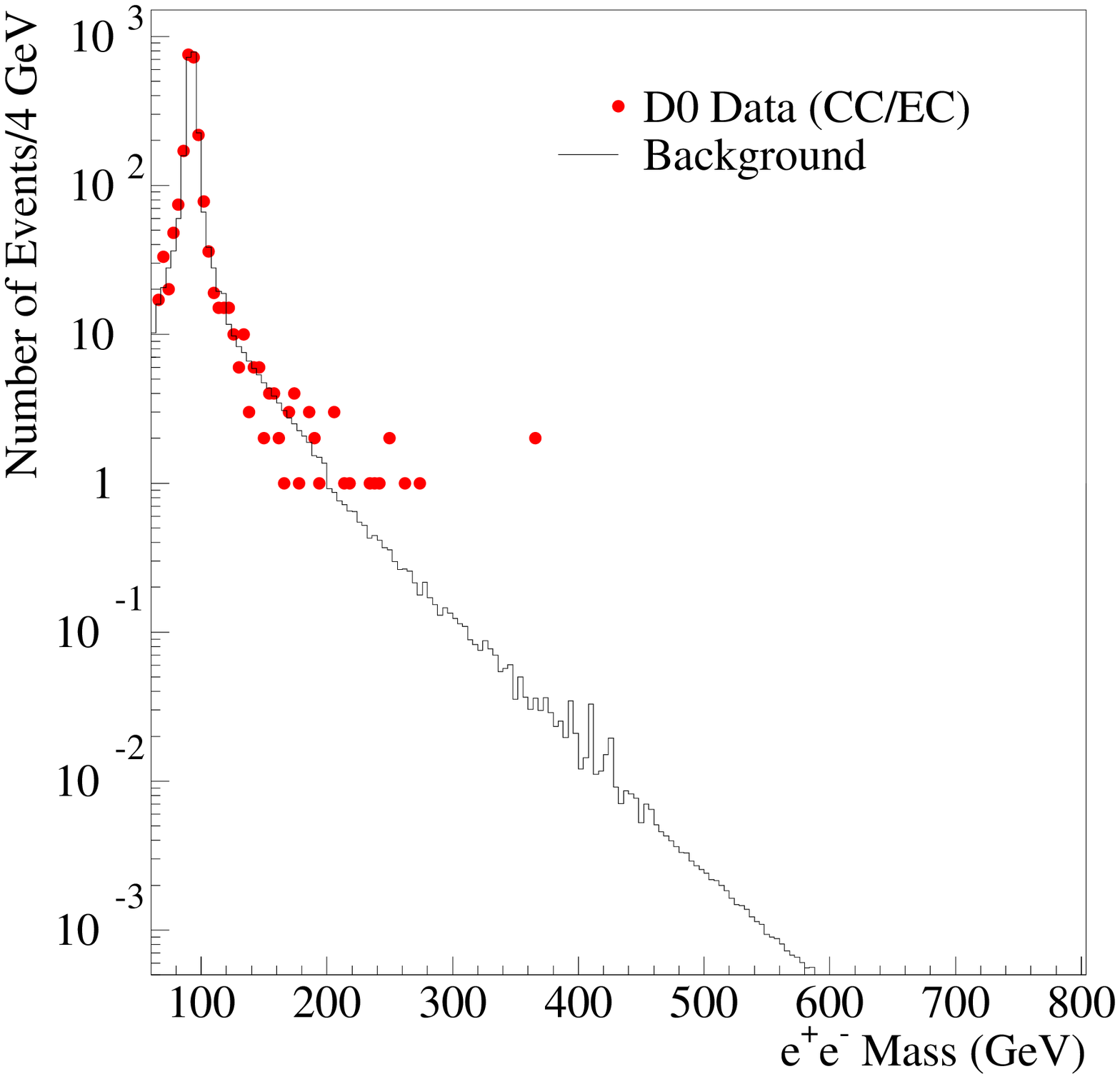,width=3.5in}}
\caption{Dielectron invariant mass spectrum for CC/EC events.}
\label{figtwo}
\end{figure}

The dielectron invariant mass spectra for events
with both electrons in the central region  (CC/CC) and with one electron
in the central region and the other in the forward region (CC/EC) are shown in 
Figs. \ref{figone} and \ref{figtwo}. The data correspond to an integrated 
luminosity of 124.8$\pm$5.1 pb$^{-1}$, taken at $\sqrt{s}$=1.8 TeV. 
The superimposed histograms represent the estimated spectrum from standard 
model processes and instrumental effects. This is dominated by two sources:
\begin{itemize}
\item  Drell-Yan process (via intermediate $\gamma^*$ and $Z^*$)
\item  Jets misidentified as electrons. This includes contributions from:
\begin{itemize}
\item Dijet events in which both jets are misidentified as electrons
\item $W(\to e\nu)$+jets events in which one of the jets is misidentified as
    an electron
\item $\gamma$+jets events in which a jet and the photon are
    misidentified as electrons
\end{itemize}
\end{itemize}
Other processes ($W\gamma$, $Z\gamma$, $t\overline t$, $WW$, and 
$\gamma^*/Z\to\tau\tau$), that can in principle also contribute to 
dielectron final states, have not been included in this analysis because these 
are at least an order of magnitude smaller than the two main backgrounds, as 
shown in Ref. \cite{Gupta}.

The Drell-Yan spectrum is estimated using the \textsc{pythia} Monte Carlo 
generator \cite{PYTHIA}. A $K$-factor is applied, as a function of  
dielectron mass, in order to normalize the cross sections from \textsc{pythia} 
to next-to-leading-order calculations\cite{vanNeerven}, 
as described in Ref. \cite{Gupta}. The uncertainty in the $K$-factor is 5\%.


The efficiencies for identification of electron-positron pairs are\cite{Gupta}:
\begin{eqnarray}
\epsilon & = & 0.814 \pm 0.014\ \mbox{for CC/CC events}; \nonumber \\
\epsilon & = & 0.479 \pm 0.010\ \mbox{for CC/EC events.}
\end{eqnarray}
The acceptance for an $e^+ e^-$-resonance signal is about 50\%, roughly 
independent of dielectron mass. The larger the dielectron mass, the larger 
is the fraction of CC/CC events, and thus the larger the total
overall efficiency. This efficiency varies between
30\% (at a mass of 140 GeV) and 40\% (at a mass of 450 GeV).
The apparent width of the resonance (dominated by the 
detector resolution) increases with the mass of the particle.


In Table \ref{tab:data-MC-comp}, we compare the observed number of events with 
standard model expectations. There is no significant excess in cross section, 
nor do we see any significant accumulation of events at one mass value, as 
expected for the decay of a narrow resonance. In the absence of a signal, we 
set an upper limit on the product of the cross section and branching fraction 
as a function of dielectron invariant mass. 

\begin{table}[th]
\caption{Comparison of observed and expected number of events, for combined 
CC/CC and CC/EC samples.}
\begin{tabular}{lrr}
\label{tab:data-MC-comp}
mass region & expected & observed \\
\hline
$>100$ GeV & 609$\pm$73 & 571 \\ 
$>200$ GeV & 26$\pm$3.4  &  32 \\
$>300$ GeV & 4.7$\pm$0.6 &   6 \\
$>400$ GeV & 1.1$\pm$0.1 &   0 \\
\end{tabular}
\end{table}

\begin {figure}[htp]
\centerline{\psfig{figure=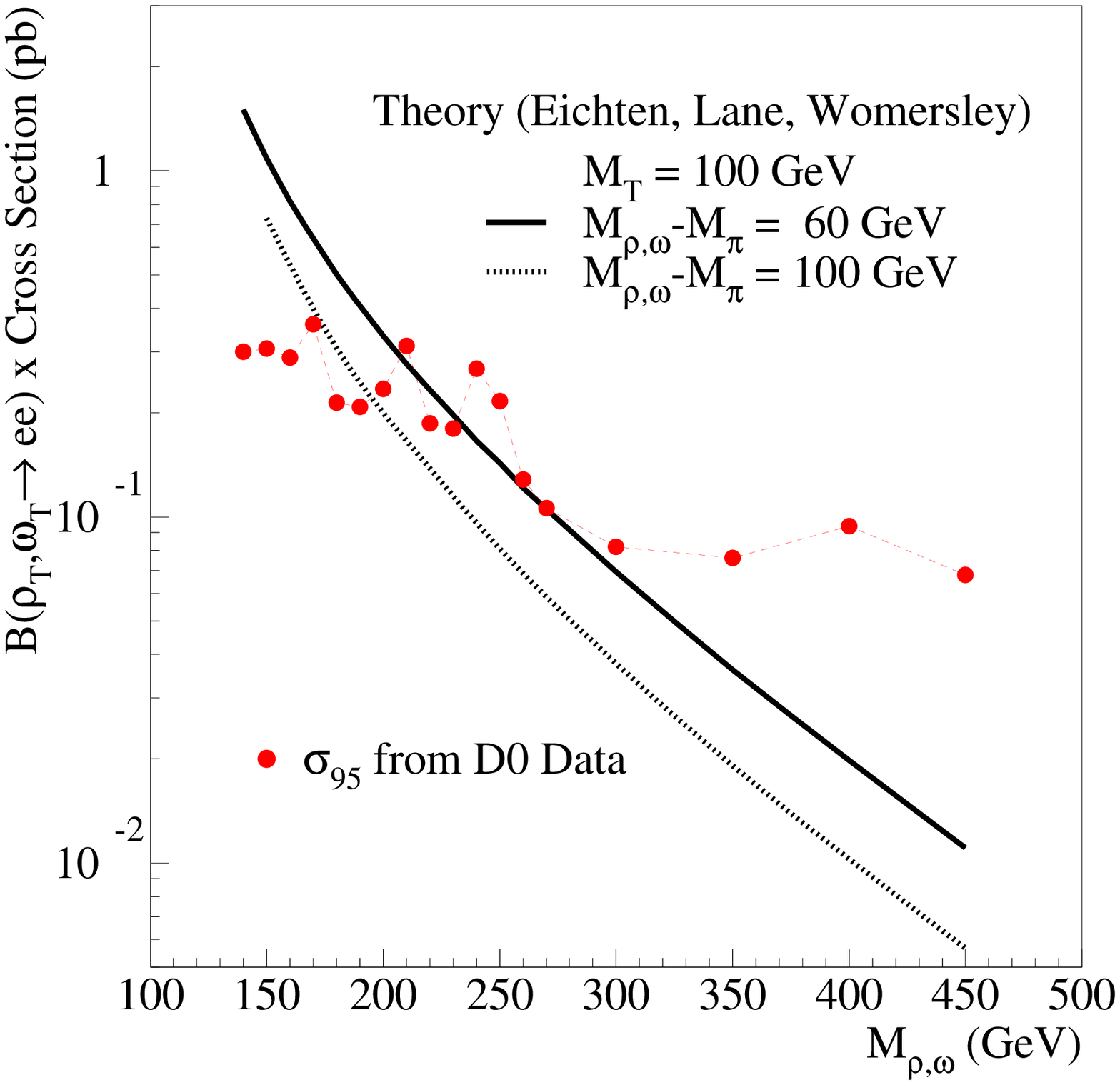,width=3.5in}}
\caption{Experimental upper limits at 95\% confidence level for $\rho_T,
\omega_T\rightarrow e^+e^-$ production compared with predictions from 
Refs.~[9,11].
$M_{\rho,\omega}$ and $M_\pi$ denote technihadron masses. }
\label{figeight}

\centerline{\psfig {figure=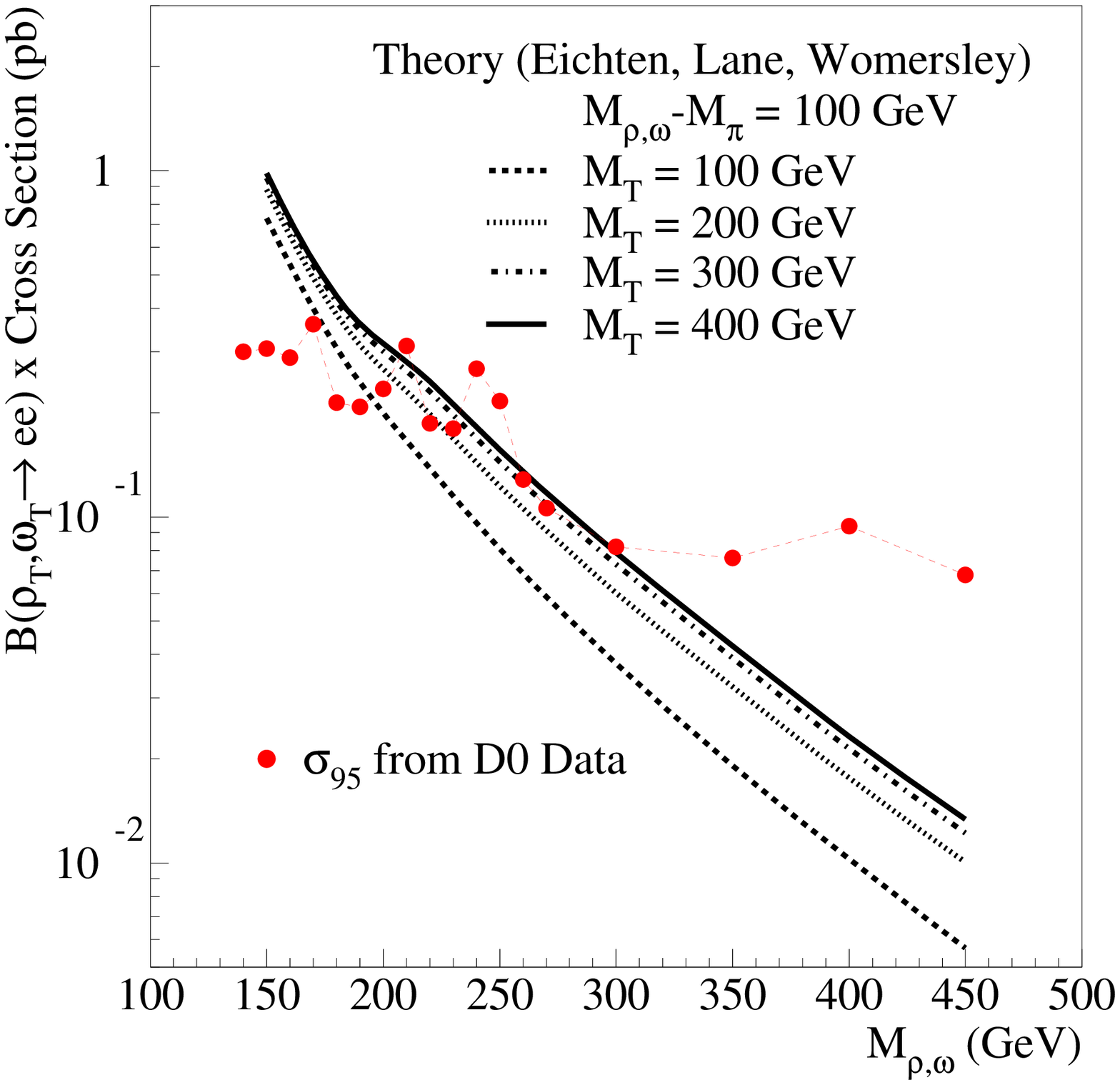,width=3.5in}}
\caption{Experimental upper limits at 95\% confidence level for $\rho_T,
\omega_T\rightarrow e^+e^-$ production compared with predictions from 
Refs.~[9,11]. $M_{\rho,\omega}$ and $M_\pi$ denote technihadron 
masses.}
\label{fignine}
\end{figure}

\begin {figure}[htb]
\centerline{\psfig {figure=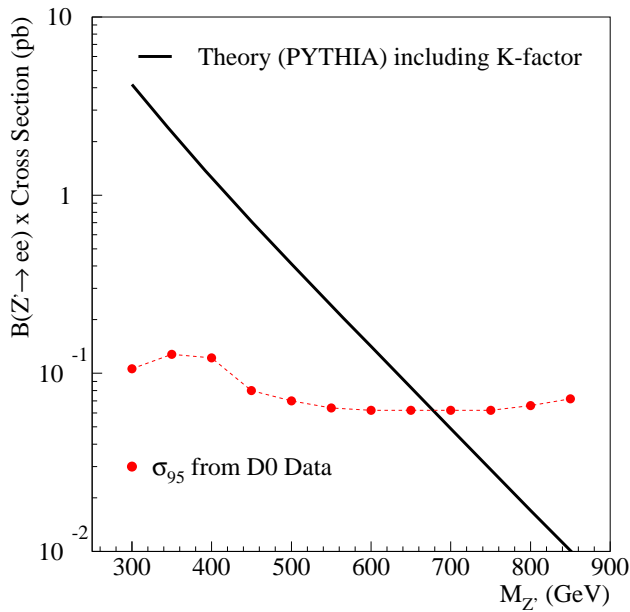,width=3.5in}}
\caption{Experimental upper limit at 95\% confidence level for 
$Z'\rightarrow e^+e^-$ production compared with predictions.}
\label{figten}
\end{figure}

We calculate the limit in a way similar to that described in Ref. 
\cite{Bertram}. We bin the spectra shown in Figs. \ref{figone} and 
\ref{figtwo} in 4 GeV wide bins. In bin $i$, we expect to see $\mu_i$ events, 
where 
\begin{equation}
\mu_i=f_i\times\sigma\times\epsilon\times{\cal L} + b^1_i +
b^2_i\times{\cal L}. \label{eq:mu}
\end{equation}
Here $f_i$ is the signal acceptance for bin $i$, $\sigma$ is the signal cross 
section multiplied by the branching fraction into $e^+e^-$, $\epsilon$ is the 
signal efficiency, ${\cal L}$ is the integrated luminosity, $b^1_i$ is the 
expected number of events with misidentified jets in bin $i$, and $b^2_i$ is 
the Drell-Yan cross section, corrected for acceptance and efficiency, 
integrated over bin $i$. The acceptance $f_i$ depends somewhat on the process 
under consideration (but not the detailed model parameters), and has been 
evaluated using Monte Carlo simulations for the specific final states 
considered below. The only unknown of these parameters is $\sigma$. We use 
Poisson statistics to calculate the probability $p_i(n_i|\mu_i)$ to see the 
$n_i$ events observed in the data given the expected value $\mu_i$. To 
account for the uncertainties in the values of the parameters that determine 
$\mu_i$, we average this probability over prior distributions for the 
parameters. The joint probability for all bins, as a function of $\sigma$, 
is then
\begin{equation}
P(\sigma) = \int\!\int\! G_{\cal L}G_\epsilon \prod_{i=1}^n\int\!\int\! 
G_{b^1_i} G_{b^2_i} p_i(n_i|\mu_i) db^2_i db^1_i d\epsilon d{\cal L} .
\end{equation}
The priors $G$ are Gaussians with means equal to the most probable parameter 
values and variances given by the square of the uncertainties. 
We calculate this probability for the CC/CC data sample ($P_{CC}(\sigma)$) 
and for the CC/EC data sample ($P_{EC}(\sigma)$) separately. 
We determine a Bayesian 95\% confidence level upper limit on the product of 
the signal cross section and branching fraction ($\sigma_{95}$) from the 
definition:
\begin {equation}
{\int_0^{\sigma_{95}} P_{CC}(\sigma)*P_{EC}(\sigma) d \sigma\over 
\int_0^{\infty} P_{CC}(\sigma)*P_{EC}(\sigma) d \sigma} = 0.95.
\end{equation}
This definition does not account for correlations in the uncertainties 
between the CC/CC and CC/EC samples because their effect on the limit is 
negligible. The resulting limits are represented by the data points in 
Figs.~\ref{figeight} and \ref{fignine} for $\rho_T$ and $\omega_T$ and in 
Fig.~\ref{figten} for $Z'$.


Topcolor-assisted technicolor models with walking gauge coupling \cite{TC2}
predict the existence of many technihadron states. The lightest of these 
technihadrons are the scalar mesons, technipions ($\pi_T^\pm$ and $\pi_T^0$), 
and the vector mesons ($\rho_T$ and $\omega_T$). These are bound states of the 
members of the lightest technifermion doublet, $U$ and $D$. They are expected 
to be produced
with substantial rates at the Fermilab Tevatron \cite{ELWone}. The vector 
mesons decay to $\gamma\pi_T$, $W\pi_T$, or fermion-antifermion pairs. 
No large isospin-violating technicolor interactions are needed to explain the 
mass difference between the top and bottom quarks. Therefore, the $\rho_T$ and 
$\omega_T$ states can be (and are assumed to be) degenerate in mass. 
As shown in Ref. \cite{Lane1}, most of the rate to dilepton final states 
originates from $\omega_T$ decays, so that our conclusions for the mass of 
the $\omega_T$ do not depend strongly on this assumption. 

The predicted products of cross sections and branching fractions for the 
processes $p\overline p\to\rho_T,\omega_T$, followed by $\rho_T,\omega_T \to 
\ell^+\ell^-$ depend on the masses of 
$\rho_T$ ($M_{\rho}$) and $\omega_T$ ($M_{\omega}$) and the mass difference 
between the vector mesons ($\rho_T$, $\omega_T$) and the 
technipions. The latter determines the spectrum of accessible decay channels.
In addition, the $\omega_T$ production cross section is sensitive to the
charges of the technifermions (taken to be $Q_U=Q_D-1=4/3$), as well as to a 
mass parameter $M_T$ that controls the rate for 
$\omega_T\rightarrow\gamma+\pi_T^0$ \cite{footnote}. The value of this
mass parameter is unknown. Scaling from the QCD decay 
$\omega\rightarrow\gamma+\pi^0$, Ref. \cite{Lane1} suggest 
a value of several hundred GeV. For all other parameters, we use the default 
values quoted in Table 2 of Ref. \cite{Lane2}.

We use recently updated calculations for the processes 
$p\overline p\to\rho_T\to\ell^+\ell^-$ and 
$p\overline p\to\omega_T\to\ell^+\ell^-$, and include a $K$-factor of 1{.}3. 
Previously published searches for technicolor particles \cite{CDF-tc} 
use an older calculation that predicted larger branching fractions for the 
dilepton decay modes. When comparing limits, this must be taken into account.
Two predictions \cite{Lane1,Lane2} for the product of cross section and 
branching fraction for the process 
$p\overline p\to(\rho_T\ \mbox{or}\ \omega_T)\rightarrow e^+ e^-$ are plotted 
in Fig.~\ref{figeight}. The two predictions shown differ in 
the assumed mass difference between the vector and scalar mesons. 
For a mass difference smaller than the mass of the $W$ boson (e.g., 60~GeV), 
the decay $\rho_T\rightarrow W+\pi_T$ is forbidden and the branching ratio 
to dielectrons is enhanced compared to the case of a mass difference of 
100~GeV, for which the $W\pi_T$ mode is allowed.
We rule out $\rho_T$ and $\omega_T$ with masses below 207~GeV,
if the mass difference between $\rho_T$ and $\pi_T^\pm$ is smaller
than the $W$-boson mass.

The limit depends on the choice of the parameter $M_T$, as illustrated in 
Fig. \ref{fignine}. In this figure, the experimental limit is compared to 
predictions in 
which the parameter $M_T$, which controls the $\omega_T$ decay rate, is 
varied. For sufficiently large values of $M_T$ ($M_T > $ 200~GeV) we can rule 
out the existence of $\rho_T$ and $\omega_T$ with masses below 203~GeV, even 
when the competing $W\pi_T$ decay mode of the technirho is open.


There is no unique prediction for the couplings of a heavy neutral gauge 
boson ($Z'$) to fermions. We assume as a benchmark that the $Z'$ has the same 
couplings to fermions as the $Z$ boson of the standard model. Thus, the width 
of the $Z'$ scales proportional to $M_{Z'}$.
We determine the product of the cross section and branching ratio using 
\textsc{pythia} and adjust for the $K$-factor \cite{D-Y K-factor}.

We set an upper limit on the product of the cross section and branching 
fraction using the same algorithm as for the technicolor particles. 
Figure \ref{figten} shows the experimental limit together with the theoretical 
cross section. For the assumed couplings, we exclude the existence of a $Z'$ 
boson below a mass of 670~GeV at the 95\% confidence level. The previous 
search by D\O\ \cite{Eppley}, using a smaller data sample, set a lower limit 
at 490~GeV. A search by CDF in both the dielectron and dimuon channels 
\cite{CDF-Zp} set a lower limit at 690~GeV. 

To summarize, based on 124.8 pb$^{-1}$ of data collected by the D\O\ detector 
at the Fermilab Tevatron during 1992--1996, we set new limits on the 
production of technirho ($\rho_T$), techniomega ($\omega_T$), and $Z'$ 
particles in \ppbar\ collisions using their decays to $e^+ e^-$.
The 95\% C.L. lower limits on the particle masses are 207~GeV for $\rho_T$ and 
$\omega_T$ states, assuming that they have equal mass and that the decay 
$\rho_T\to\pi_T+W$ is kinematically forbidden, and 
670~GeV for $Z'$ bosons with standard model couplings to fermions.

We thank K.~Lane for helpful discussions and S.~Mrenna for adding the 
technicolor model to \textsc{pythia}. We also acknowledge the hospitality of 
the Aspen Center for Physics, where this analysis was initiated. 

We thank the staffs at Fermilab and collaborating institutions, 
and acknowledge support from the 
Department of Energy and National Science Foundation (USA),  
Commissariat  \` a L'Energie Atomique and 
CNRS/Institut National de Physique Nucl\'eaire et 
de Physique des Particules (France), 
Ministry for Science and Technology and Ministry for Atomic Energy (Russia),
CAPES and CNPq (Brazil),
Departments of Atomic Energy and Science and Education (India),
Colciencias (Colombia),
CONACyT (Mexico),
Ministry of Education and KOSEF (Korea),
CONICET and UBACyT (Argentina),
The Foundation for Fundamental Research on Matter (The Netherlands),
PPARC (United Kingdom),
Ministry of Education (Czech Republic),
and the A.P.~Sloan Foundation.


\begin{references}


\bibitem{D0NIM}
S.~Abachi {\sl et al.} (D\O\ Collaboration), 
Nucl. Instrum. Methods Phys. Res., A {\bf 338}, 185 (1994).

\bibitem{Gupta}
B.~Abbott {\sl et al.} (D\O\ Collaboration), 
Phys. Rev. Lett. {\bf 82}, 4769 (1999).

\bibitem{electron} The D\O\ detector does not determine the sign of the electric charge of electrons and positrons. We therefore use the generic term ``electron'' to refer to both $e^-$ and $e^+$.

\bibitem{PYTHIA} T. Sj\"ostrand, Comput. Phys. Commun. {\bf82}, 74 (1994).

\bibitem{vanNeerven} R.~Hamberg, W.~L.~van~Neerven, and T.~Matsuura, Nucl. Phys. {\bf B359}, 343 (1991).

\bibitem{Bertram} 
B.~Abbott {\sl et al.} (D\O\ Collaboration), Phys. Rev. Lett. {\bf 82}, 2457 (1999).

\bibitem{TC2} C. Hill, Phys. Lett. B {\bf 345}, 483 (1995); K. Lane, Phys. Rev. D {\bf 54}, 2204 (1996) and references therein.

\bibitem{ELWone}
E. Eichten, K. Lane and J. Womersley, Phys. Lett. B {\bf 405}, 305 (1997).

\bibitem{Lane1}
K. Lane, Phys. Rev. D {\bf60}, 075007 (1999).

\bibitem{footnote}
In Ref. \cite{Lane1}, two parameters, $M_A$ for axial-vector and 
$M_V$ for vector couplings, appear. Their values are expected to be 
comparable. We set $M_A=M_V=M_T$.

\bibitem{Lane2}
K. Lane, ``Technihadron production and decay rates in the technicolor straw 
man model'', hep-ph/9903372.

\bibitem{CDF-tc}
T.~Affolder {\sl et al.} (CDF Collaboration), Phys. Rev. Lett. {\bf84}, 1110 (2000).

\bibitem{D-Y K-factor}
R.~Hamberg, W.L.~van Neerven, T.~Matsuura, Nucl. Phys. B {\bf359}, 343 (1991). 

\bibitem{Eppley} 
S.~Abachi {\sl et al.} (D\O\ Collaboration), Phys. Lett. B {\bf 385}, 471 (1996).

\bibitem{CDF-Zp}
F.~Abe {\sl et al.} (CDF Collaboration), Phys. Rev. Lett. {\bf79}, 2192 (1997).

\end{references}
\end{document}